\documentstyle[12pt]{article}

\begin{document}

\vspace*{0.25in}
\begin{large}
\centerline{\bf An Extended Liouville Equation}
\centerline{\bf for Variable Particle Number Systems}
\end{large}
\vskip 0.5in

\noindent {\it Michael H. Peters}$^{a)}$
\newline\noindent {\it Department of Chemical Engineering, Florida State University} 
\newline\noindent {\it and Florida A$\&$M University, Tallahassee, FL  32310}
\newline\noindent (Submitted September 1998)

\vskip 0.25in

\noindent It is well-known that the Liouville equation of statistical mechanics is restricted to systems where the total number of particles (N) is fixed.  In this paper, we show how the Liouville equation can be extended to systems where the number of particles can vary, such as in open systems or in systems where particles can be annihilated or created.  A general conservation equation for an arbitrary dynamical variable is derived from the extended Liouville equation following Irving and Kirkwood's$^{2}$ technique.  From the general conservation equation, the particle number conservation equation is obtained that includes general terms for the annihilation or creation of particles. It is also shown that the grand canonical ensemble distribution function is a particular stationary solution of the extended Liouville equation, as required.  In general, the extended Liouville equation can be used to study nonequilibrium systems where the total number of particles can vary.

\vfill

\noindent $^{a)}$ e-mail: peters@eng.fsu.edu

\newpage

\noindent {\bf I. Introduction}

Modern molecular theories of equilibrium thermodynamics and nonequilibrium thermodynamics are based almost entirely on the Liouville equation which describes the conservation of an ensemble of phase points as they move through phase space according to a specified Hamiltonian$^{1}$.  Unfortunately, the Liouville equation is based on a fixed number of particles and cannot describe systems where the total number of particles varies.  For example, the Liouville equation cannot describe the statistical mechanical, grand canonical ensemble used in the development of thermodynamic properties of open, equilibrium systems.  In this letter, we extend the Liouville equation to systems where the number of particles varies.  To our knowledge this has not been done before and has restricted the theoretical development of nonequilibrium thermodynamics.  These systems include open systems, where the particle number can vary because of open boundaries, or closed or open systems where particles can be annihilated or created either classically by chemical reactions or due to quantum mechanical effects. 

\noindent {\bf II. The Extended Liouville Equation}

The Liouville equation describes the behavior of a collection or ensemble of phase points as they move through a multidimensional space, or phase-space.  Each phase point represents the position (or generalized coordinates $q_{i}$) and momentum (or conjugate momentum $p_{i}$) of all $N$ molecules or particles in the system.  The phase points tend to be concentrated in regions of phase space where it is most likely to find the $N$ particles with a certain momentum and position.  Thus, the Liouville density function, $\rho_L$, can be interpreted (aside from a normalization constant) as a probability density function, i.e., $\rho_{L} d{\bf q}^{N}d{\bf p}^{N}$ is proportional to the probability of finding a phase point in a multidimensional region between $({\bf q}^{N}, {\bf p}^{N})$ and $({\bf q}^{N} + d{\bf q}^{N}, {\bf p}^{N} + d{\bf p}^{N})$ at any time, $t$.  Here and throughout we use the short-hand notation ${\bf q}^{N}=q_{1}, \cdots, q_{3N}$ and ${\bf p}^{N}={p}_{1}, \cdots, p_{3N}$.  It can be readily shown that the Liouville density function obeys the conservation equation$^{1}$

$$\frac{\partial \rho_{L} ({\bf p}^{N}, {\bf q}^{N}, t)}{\partial t} + \sum_{i=1}^{3N} \left[\frac{\partial \rho_{L}}{\partial q_{i}}\frac{\partial H}{\partial p_{i}}-\frac{\partial \rho_{L}}{\partial p_{i}}\frac{\partial H}{\partial q_{i}}\right] = 0 \eqno(1)$$

\noindent where $H({\bf p}^{N}, {\bf q}^{N}, t)$ is the Hamiltonian or total energy of the $N$-particle system.

It is to be noted that the Liouville equation describes the phase space behavior of the density function $\rho_{L}({\bf p}^{N}, {\bf q}^{N}, t)$ under the conditions where the number of particles $N$ is a constant.  To extend the Liouville equation to a system where the number of particles or dimensionality can vary, consider the extended space $({\bf p}^{N}, {\bf q}^{N}, N, t)$ shown in Fig.1.  The total members of the ensemble in extended space is a constant denoted by $M$, and the density of phase points is denoted by $\rho_{N}({\bf p}^{N}, {\bf q}^{N}, N, t)$.  Thus, we have that 

$$\sum_{N}\int \int \rho_{N}({\bf p}^{N}, {\bf q}^{N}, N, t)d{\bf p}^{N}d{\bf q}^{N}=M \eqno(2)$$

\noindent As shown in Fig.1, the extended space can be divided into ``canonical" hyperplanes with the number of phase points at time $t$ in each hyperplane denoted by $M_{N_1}$, $M_{N_2}$, etc., where

$$M=\sum_{N_i}M_{N_i} \eqno(3)$$

The phase points in each hyperplane have the same number of particles or dimensionality and, hence, we call these ``canonical" hyperplanes.  As the system evolves in time, the phase points can move within the plane, if the dimensionality $(N)$ does not change, or ``jump" between planes, if the dimensionality $(N)$ of the phase point changes. In any event, the total number of phase points $M$ remains constant with time.

Now, it is not possible to straightforwardly extend the Liouville equation by considering differential changes in ${\bf q}^{N}$, ${\bf p}^{N}$ and $N$ (the extended space), since as $N$ changes so does the dimensionality of the phase space coordinates $({\bf q}^{N}, {\bf p}^{N})$.  However, if we define a new function

$$\bar{\rho}(N, t)=
\int\int \rho_{N}
({\bf p}^{N}, {\bf q}^{N}, N, t)
d{\bf p}^{N}d{\bf q}^{N}=M_{N}\eqno(4)$$

\noindent then, in order to conserve the total number of phase points $M$, it can be readily shown (Appendix A) that $\bar{\rho}$ must obey the one-dimensional conservation equation

$$\frac{\partial \bar{\rho}}{\partial t}+\frac{\partial}{\partial N}[\bar{\rho}v(N, t)]=0 \eqno(5)$$

\noindent where we have introduced a ``velocity" function, $v(N, t)$,

$$v(N, t)\equiv v^{+} (N, t) - v^{-}(N, t)\eqno(6)$$

\noindent 

\noindent which physically represents the net fractional number of phase points leaving the $N^{th}$ hyperplane per unit time (Appendix A); $v^{+}$ represents the fraction of phase points leaving the $N^{th}$ hyperplane to the $(N+1)^{th}$ hyperplane, and $v^{-}$ are the fraction per unit time leaving to the $(N-1)^{th}$ hyperplane.  Now, within each hyperplane we introduce the Liouville density $\rho_{L}({\bf p}^{N}, {\bf q}^{N}, t; N)$, defined as a conditional density function

$$\rho_{L}({\bf p}^{N}, {\bf q}^{N}, t; N)\equiv 
\frac{\rho_{N}({\bf p}^{N}, {\bf q}^{N}, N, t)}{\int \rho_{N}d{\bf p}^{N}d{\bf q}^{N}} $$

$$=\frac{\rho_{N}}{\bar{\rho}} \eqno(7)$$

\noindent The phase points that move within the hyperplane over a differential time, $dt$, obey the usual, Liouville equation, Eq.(1),

$$\frac{\partial \rho_{L}}{\partial t}+\sum_{i=1}^{3N}
\left(
\frac{\partial \rho_{L}}{\partial q_{i}}\frac{\partial H}{\partial p_{i}}-\frac{\partial \rho_{L}}{\partial p_{i}}\frac{\partial H}{\partial q_{i}} 
\right)=0 
\eqno(8)$$

Thus, the space-time behavior of the extended Liouville density function $\rho_{N}({\bf p}^{N}, {\bf q}^{N}, N, t)$ is uniquely determined through Eqs.(5), (7) and (8).  Physically, we can envision that phase points move in each hyperplane over a differential time, $dt$, according to Eq.(8).  At the end of the differential time, some phase points ``jump" into new canonical hyperplanes according to Eq.(5).  Thus, the number of phase points in each canonical hyperplane with a fixed dimensionality continually change with time.  Although both the dimensionality change and time change are discrete, the continuous representation given by Eq.(5) should approximately hold for $N$ large and $dt \rightarrow 0$.

Multiplying Eq.(5) by $\rho_{L}$, Eq.(8) by $\bar{\rho}$, and combining using Eq.(7) we can write {\it an extended Liouville equation} as 

$$\frac{\partial \rho_{N}\left({\bf q}^{N}, {\bf p}^{N}, N, t \right)}
{\partial t} 
+ \bar{\rho} (N, t) \left\{
\sum_{i}\frac{\partial \rho_{L}}{\partial q_{i}}
\frac{\partial H}{\partial p_{i}}
-\frac{\partial \rho_{L}}{\partial p_{i}}\frac{\partial H}{\partial q_{i}} \right\}$$
 
$$+ \rho_{L}({\bf q}^{N}, {\bf p}^{N}, t; N)\frac{\partial}{\partial N}[ \bar{\rho}v(N, t)]=0 \eqno(9)$$

\noindent or,

$$\frac{\partial \ln \rho_{N}}{\partial t} + \sum_{i}
\frac{}{}
\left.(
\frac{\partial \ln \rho_{L}}
{\partial q_{i}}
\frac{\partial H}
{\partial p_{i}} -
\frac{\partial \ln \rho_{L}}
{\partial p_{i}}
\frac{\partial H}
{\partial q_{i}}
\frac{}{}\right.)
$$

$$ + v(N, t)\frac{\partial \ln \bar{\rho}}{\partial N}+ \frac{\partial v}{\partial N} =0  \eqno(10)$$

In the next section, we derive a general conservation (transport) equation for an arbitrary dynamic variable where the total number of particles can vary.  From this we obtain a particle number conservation equation which includes a general term accounting for the annihilation or creation of particles.

\noindent{\bf  III. General Conservation Equation for Variable Particle Number Systems}

In Cartesian coordinates, the extended Liouville equation can be written 

$$\frac{\partial \rho_{N}({\bf r}^{N}, {\bf p}^{N}, N, t)}{\partial t} + \bar{\rho}(N, t)\left\{ \sum_{i} \left[ \frac{\partial}{\partial {\bf r}_{i}} \cdot (\frac{{\bf p}_{i}}{m} \rho_{L}) + \frac{\partial }{\partial {\bf p}_{i}} \cdot ({\bf F}_{i}\rho_{L})\right]\right\}$$

$$ + \rho_{L}({\bf r}^{N}, {\bf p}^{N}, t; N)\frac{\partial}
{\partial N}[\bar{\rho}v(N, t)]=0 \eqno(11)$$

\noindent where we will now normalize the density function $\bar{\rho}$ by the total ensemble number $M$ (fixed), i.e., 

$$\sum_{N}\bar{\rho}=1 \eqno(12)$$

$$\int \int \rho_{L}d{\bf r}^{N}d{\bf p}^{N}=1 \eqno(13)$$

$$\sum_{N} \int \int \rho_{N}d{\bf r}^{N}d{\bf p}^{N} = 1 \eqno(14)$$

Extending Irving and Kirkwood's formalism$^{2}$, the ``conservation equations" for variable particle number systems can be obtained by first considering an arbitrary quantity $\alpha({\bf r}^{N}, {\bf p}^{N}, N)$ that does not depend explicitly on time.  The average or expectation value of $\alpha$ is defined as

$$<\alpha> \equiv \sum_{N} \int \int \alpha ({\bf r}^{N}, {\bf p}^{N}, N)
\rho_{N} d{\bf r}^{N} d{\bf p}^{N} \eqno(15)$$

Now, multiplying the extended Liouville equation by $\alpha$ and integrating over all $({\bf r}^{N}, {\bf p}^{N})$ and summing over all $N$ yields

$$\frac{\partial <\alpha>}{\partial t} - \sum_{N} \bar{\rho} 
\int \int \sum_{i} \rho_{L}
\frac{{\bf p}_{i}}{m} \cdot \frac{\partial \alpha}{\partial {\bf r}_{i}} d{\bf r}^{N}d{\bf p}^{N}$$ 

$$- \sum_{N} \bar{\rho} \int \int  \sum_{i} \rho_{L}
{\bf F}_{i} \cdot \frac{\partial \alpha}{\partial {\bf p}_{i}}  d{\bf r}^{N}d{\bf p}^{N}$$ 

$$+ \sum_{N} \frac{\partial}{\partial N}(\bar{\rho}v) \int \int \alpha 
\rho_{L} d{\bf r}^{N}d{\bf p}^{N} \eqno(16)$$ 

\noindent where we have used the conditions

$$\left.
\begin{array}{rcr}
\displaystyle\rho_{L} & \displaystyle \rightarrow  & 0\\
\displaystyle {\bf p}_{i}\rho_{L} & \displaystyle \rightarrow & 0\\
\displaystyle \alpha {\bf p}_{i}\rho_{L} & \displaystyle \rightarrow & 0
\end{array} \displaystyle \right\}\;as \;{\bf r}_{i}, {\bf p}_{i} \;\rightarrow \; \infty \eqno(17)$$ 

\noindent Equation (16) is the general conservation equation for $\alpha$.

\noindent {\it Particle Number Conservation}

Again, extending Irving and Kirkwood's formalism$^{2}$ for variable particle number systems, the equation for particle number conservation (ordinary number density) can be obtained from Eq.(16) by setting

$$\alpha ({\bf r}_{N}, {\bf p}_{N}, N)= \sum_{k=1}^{N} \delta ({\bf r}_{\kappa}-{\bf r}) \eqno(18)$$

\noindent Thus, 

$$<\alpha> = n({\bf r}, t) \equiv 
\sum_{N} \bar{\rho} \int \int \left[ \sum_{k=1}^{N} \delta ({\bf r}_{k} - {\bf r})\right] \rho_{L} d{\bf r}^{N}d {\bf p}^{N}$$

$$= \sum_{N} \bar{\rho} N \int \int \rho_{1}({\bf r}, {\bf p}, t; N) d{\bf p}$$

$$= \sum_{N} \bar{\rho} N \psi_{1} ({\bf r}, t; N) \eqno(19)$$

\noindent where $\psi_{1}({\bf r}, t; N)d{\bf r}$ is the conditional probability of finding a particle between {\bf r} and {\bf r} + d{\bf r}, given a system of $N$ total particles at time $t$, i.e.,

$$\psi_{1}({\bf r}, t; N)\equiv \int \rho_{1} ({\bf r}, {\bf p}, t; N)d{\bf p}\eqno(20)$$

\noindent Note that for a system at equilibrium 

$$\psi_{1}=\frac{1}{V}\quad, \quad{\rm equilibrium}  \eqno(21)$$

\noindent where $V$ is the volume of the system; thus

$$n({\bf r}, t) = \frac{1}{V} \sum_{N}\bar{\rho}N = \frac{\bar{N}}{V}\quad, \quad {\rm equilibrium} \eqno(22)$$

\noindent where $\bar{N}$ is the equilibrium average number of particles in the volume $V$.

Analyzing the remaining terms in Eq.(16)

$$-\sum_{N} \bar{\rho} \int \int \left[ \sum_{i} \rho_{L}
\frac{{\bf p}_{i}}{m}\cdot \frac{\partial \alpha}{\partial {\bf r}_{i}}     \right] d{\bf r}^{N}d{\bf p}^{N} $$

$$
= \frac{\partial}{\partial {\bf r}} \cdot \left[ \sum_{N} \bar{\rho} N \int \rho_{1} \frac{\bf p}{m} d{\bf p}\right]$$

$$
= \frac{\partial }{\partial {\bf r}} \cdot [ n ({\bf r}, t) {\bf v}_{0}({\bf r}, t)]
\eqno(23)$$

\noindent where we have defined the number average velocity as

$${\bf v}_{0}({\bf r}, t)\equiv \frac{\sum_{N} \bar{\rho} N \int \rho_{1}\frac{\bf p}{m}d{\bf p}}{\sum_{N}\bar{\rho}N\psi_{1}}\eqno(24)$$

\noindent and, for the final term,

$$\sum_{N} \left[\frac{\partial }{\partial N}(\bar{\rho}v)\right]
\int \int \alpha \rho_{L} d{\bf r}^{N} d{\bf p}^{N}$$

$$=\sum_{N} \left[\frac{\partial }{\partial N} (\bar{\rho}v)\right]
N\psi_{1}({\bf r}, t; N) \eqno(25)$$

\noindent Summarizing, the particle number conservation equation becomes

$$\frac{\partial n({\bf r}, t)}{\partial t} + \frac{\partial }{\partial {\bf r}} \cdot (n {\bf v}_{0}) + \sum_{N} \left[ \frac{\partial }{\partial N}(\bar{\rho}v)  \right]
N \psi_{1} ({\bf r}, t; N) = 0 \eqno(26)$$ 

\noindent The third term on the left-hand side accounts for the creation or annihilation of particles in the system as considered in more detail in the example below.

\noindent {\it Example.  First-Order Particle Annihilation}

As a simple example of the application of Eq.(26) consider the problem of particle annihilation where the fractional number of particles leaving the $N^{th}$ hyperplane to the $(N-1)^{th}$ hyperplane is a constant $k$.  Thus, Eq.(5) becomes

$$\frac{\partial \bar{\bf \rho}}{\partial t} - k\frac{\partial \bar{\rho}}{\partial N}=0  \eqno(27)$$

\noindent or, in ``finite $N$" form (Appendix A)

$$
\frac{\partial \bar{\rho}(N, t)}{\partial t} - k [\bar{\rho}(N+1, t) - \bar{\rho}(N, t)]=0  \eqno(28)
$$ 

\noindent Consider that initially all phase points are in the $N_{0}$ hyperplane, i.e., 

$$\bar{\rho}(N_{0}, 0)=1 \eqno(29)
$$

\noindent and

$$\bar{\rho}(N, 0)=0\quad, \quad N \in [0, N_{0}-1]  \eqno(30)$$

\noindent Around the $N_{0}$ hyperplane, the population balance equation is 

$$
\frac{\partial \bar{\rho}(N_{0}, t)}{\partial t} + k\bar{\rho} (N_{0}, t)=0\eqno(31)
$$

\noindent Thus, using the initial state Eq.(29), we obtain

$$
\bar{\rho}(N_{0}, t)=e^{-kt} \eqno(32)
$$

Now, performing a balance around the $N_{0}-1$ hyperplane using Eq.(32) leads to

$$
\frac{\partial \bar{\rho}(N_{0}-1, t)}{\partial t} + k\bar{\rho} (N_{0}-1, t)=k e^{-kt} \eqno(33)
$$

\noindent and, using the initial condition Eq.(30), we obtain

$$\bar{\rho}(N_{0}-1, t)=kte^{-kt} \eqno(34)$$

Continuing this process, for the $N^{th}$ hyperplane we must have 

$$
\bar{\rho}(N, t) = \frac{(kt)^{m}}{m!} e^{-kt} \eqno(35)
$$

\noindent where $m=N_{0}-N$, $m \in [0, N_{0}-1]$.  We note that for $N_{0} \rightarrow \infty$,

$$\sum\bar{\rho}(N, t)=e^{-kt}\sum_{m=0}^{\infty} \frac{(kt)^{m}}{m!}=1\eqno(36)
$$

\noindent as required.  Note that for small or finite values of $N_{0}$, we must consider a population balance around the $N=0$ hyperplane (total particle annihilation plane) in order to satisfy the normalization condition.

Now consider the annihilation/creation term in the particle number conservation equation, Eq.(26).  We have that 

$$\frac{\partial }{\partial N}(\bar{\rho}v)=-\frac{\partial \bar{\rho}}{\partial t}=k\left[ 1-\frac{(N_{0}-N)}{kt} \right] \bar{\rho}\eqno(37)
$$

\noindent Substituting into the third term on the l.h.s of Eq.(26) leads to the first-order decay kinetic expression, under the conditions $kt >> (N_{0}-N)$, as

$$
\frac{\partial n ({\bf r}, t)}{\partial t} + \frac{\partial }{\partial {\bf r}}\cdot(n{\bf v}_{0}) + \sum_{N}k \bar{\rho} N \psi_{1}=0\quad, \quad kt >>
(N_{0}-N) \eqno(38)
$$

\noindent or

$$
\frac{\partial n ({\bf r}, t)}{\partial t} + \frac{\partial }{\partial {\bf r}}\cdot(n{\bf v}_{0}) + kn=0\quad, \quad kt >>
(N_{0}-N) \eqno(39)
$$ 

\noindent Note that for small times compared to $(N_{0}-N)/k$ the decay kinetics are not first-order, even for a constant rate of phase point hopping, $k$.  This example helps illustrate the generality of the extended Liouville equation.  Next, we address the question as to whether the extended Liouville equation leads to an equilibrium behavior consistent with the grand canonical ensemble.

\noindent {\bf IV. Equilibrium Behavior.  Grand Canonical Ensemble}

Under equilibrium conditions, $\partial \bar{\rho}/\partial t = 0$, the general stationary solution to Eq.(5) can be written as 

$$\bar{\rho}(N, t)=C_{1}f(N) \eqno(40)$$

$$v(N, t)=C_{2}[f(N)]^{-1} \eqno(41)$$

\noindent where $C_{1}$ and $C_{2}$ are constants to be determined and $f(N)$ is an arbitrary function of $N$.  Now, it is already well-known that a {\it particular stationary solution} to Eq.(8), under the conditions $\partial \rho_{L}/\partial t=0$, can be written in dimensionally correct form as$^{3}$

$$\rho_{L}({\bf p}^{N}, {\bf q}^{N}; N)=
\frac{C_{3}}{h^{3N}N!} {\rm exp}
\left\{-\beta H ({\bf p}^{N}, {\bf q}^{N})\right\}
\eqno(42)$$

\noindent where $h$ is Planck's constant that has the units of action (length $\times$ momentum), $H({\bf p}^{N}, {\bf q}^{N})$ is the Hamiltonian or total energy, and $C_{3}$ and $\beta$ are constants to be determined.  

The constant $\beta$ can be determined by requiring that the average kinetic energy on each hyperplane be a uniform constant (canonical ensemble on the hyperplane), i.e., in Cartesian coordinates

$$\frac{(1/2m)\int p_{k}^{2}\rho_{L}({\bf r}^{N}, {\bf p}^{N})d{\bf p}^{N}}
{\int \rho_{L}d{\bf p}^{N}}=\frac{3}{2}kT \eqno(43)$$

\noindent and 

$$H({\bf r}^{N}, {\bf p}^{N})= \sum_{i=1}^{3N}\frac{1}{2}
\frac{p_{i}^{2}}{m} + \Phi ({\bf r}^{N})\eqno(44)$$

\noindent where $\Phi ({\bf r}^{N})$ is the total potential.  Thus, we obtain

$$\beta=\frac{1}{kT}\eqno(45)$$

Combining Eqs.(40) and (42) gives the extended equilibrium density function as $(C_{4}\equiv C_{1}C_{3})$

$$\rho_{N}({\bf p}^{N}, {\bf q}^{N}, N)
=\frac{C_{4}}{h^{3N}N!} {\rm exp}
\left\{-\beta H ({\bf p}^{N}, {\bf q}^{N})\right\}f(N)
\eqno(46)$$

The constant $C_{4}$ can be determined from Eq.(2) as

$$C_{4}=\frac{M}{\sum_{N}\frac{1}{h^{3N}N!} f(N)\int\int{\rm exp}\left\{-\beta H ({\bf p}^{N}, {\bf q}^{N})\right\} d{\bf p}^{N}d{\bf q}^{N}}
\eqno(47)$$

\noindent Finally, normalizing $\rho_{N}$ with the total number of phase points $M$ leads to the well-known form for the (classical) probability distribution function for the grand canonical ensemble as$^{5}$,

$$\rho_{N} ({\bf p}^{N}, {\bf q}^{N}, N)
=[E^{Grand}N!h^{3N}]^{-1} f(N){\rm exp}
\left\{-\frac{H({\bf p}^{N}, {\bf q}^{N})}{kT}\right\}\eqno(48)$$

\noindent where the (classical) grand partition function $E_{N}^{Grand}$ is given by 

$$E^{Grand} = \sum_{N}E_{N}^{can} f(N) \eqno(49)$$

\noindent where $E_{N}^{can}$ is the classical canonical partition function 

$$E_{N}^{can}=[N!h^{3N}]^{-1}\int\int {\rm exp}\frac{}{}
\left\{-\beta H({\bf p}^{N}, {\bf q}^{N})\frac{}{}\right\}d{\bf p}^{N}d{\bf q}^{N}\eqno(50)$$

Now, it is well-known from most probable distribution methods$^{6}$ that

$$
f(N)=e^{\lambda N} \eqno(51)
$$

\noindent for the equilibrium grand canonical ensemble.  All that we can state from the extended Liouville equation is that Eq.(51) is a {\it particular stationary solution} to Eq.(5)$^{4}$.  We are presented with the same difficulty as in the equilibrium solution to the ordinary Liouville equation, Eq.(29), i.e., the lack of proof of the uniqueness of Eq.(29) under equilibrium conditions - a well-known problem in statistical mechanics.

Finally, we note that to complete the grand canonical ensemble description, we must still specify $\lambda$.  This procedure is already well-known and is generally accomplished by comparison to the thermodynamic relation$^{6}$.

$$T dS= dU + pdV- \mu dN \eqno(52)$$

\noindent where $\mu$ is the chemical potential

$$\frac{}{}\mu \equiv \left(
\frac{\partial U}{\partial N}
\right)_{S,V} \eqno(53)$$

\noindent which leads to

$$\lambda = + \frac{\mu}{kT} \eqno(54)$$

\noindent and an expression for the entropy, $S$, in the grand canonical ensemble as

$$S=\frac{\bar{U}}{T}-\frac{\bar{N}\mu}{T} + k \ln E^{Grand} \eqno(55)$$

\noindent where $\bar{U}$ and $\bar{N}$ are the average internal energy and particle number over the grand canonical ensemble

$$\bar{U}=\sum_{N}
\int\int H \rho_{N}d{\bf p}^{N}d{\bf q}^{N}  \eqno(56)$$

$$\bar{N}=\sum_{N}
\int\int N \rho_{N}d{\bf p}^{N}d{\bf q}^{N}  \eqno(57)$$

The results and procedures given in Eqs.(52) - (57) are already well established, and we present them only for completeness.  

\noindent {\bf IV.  Conclusions}

The importance of the analysis given here lies in the evolution equations, Eqs.(5) - (8), or Eqs.(9) or (10), that describe, in general, the nonequilibrium behavior of the extended Liouville density $\rho_{N}({\bf p}^{N}, {\bf q}^{N}, N, t)$ for systems where the number of particles is not fixed.  A general conservation equation can be derived, following Irving and Kirkwood's paradigm$^{2}$, for systems where the total number of particles can vary.  As an example, the general particle number conservation equation was obtained that was shown to include a general term for particle annihilation or creation. 
It was also demonstrated that the grand canonical ensemble distribution is a particular stationary solution to the extended Liouville equation.  As with the equilibrium solution to the ordinary Liouville equation, {\it uniqueness} of the variable particle number solution under equilibrium conditions remains to be demonstrated.  It is to be noted that the general stationary solution to the extended Liouville equation, Eqs.(40) and (41), demonstrates that $v(N)$ is smallest where $\bar{\rho}$ is largest, i.e., near the most probable state.  Conversely, phase points move or change their dimensionality more rapidly from regions where $\bar{\rho}$ is small, i.e., near the least probable state. 
Further treatment of the extended Liouville equation in nonequilibrium systems is under current investigation.

\noindent \rule{2.5in}{.25mm}

\noindent \begin{tabular}{ll}
$^{1}$ & Liouville, {\it Journ. de Math.} {\bf 3}, 349 (1838), as cited by R.C. Tolman, {\it The Principles of}\\
& {\it Statistical Mechanics}, Oxford University Press, 1938 (p.49).\\
$^{2}$ & J.H. Irving and J.G. Kirkwood, {\it J. Chemical Phys.}, {\bf 18}, 817 (1950).\\
$^{3}$ & See J.O. Hirschfelder, C.F. Curtiss and R.B. Bird, {\it Molecular Theory of Gases and} \\
& {\it Liquids}, Wiley, NY, 1960 (Eq.2.1 - 20, p.86).\\ 
$^{4}$ & From Eq.(5), we have that $\partial \ln \bar{\rho}/\partial N = \partial \ln v/\partial N = \lambda$ which gives Eq.(51) as a particular\\
& solution.\\ 
$^{5}$ & See, e.g., Ref. 3 (Eq.2.1 - 22, p.87).\\
$^{6}$ & D.A. McQuarrie, {\it Statistical Mechanics}, Harper and Row, NY, 1976 (Ch.3, p.54).
\end{tabular}

\vskip 12pt
\noindent {\bf Acknowledgements}

I want to thank Professor I. Bitsanis for suggesting this problem to me.

\noindent {\bf Appendix A.  Conservation Equation for $\bar{\rho}(N, t)$}

Consider a simple population balance around the $N^{th}$ hyperplane (see Fig.1). Since phase points cannot be created or destroyed, we must have 

$$
\frac{\partial \bar{\rho}(N, t)}{\partial t}+ 
[\bar{\rho}(N, t)v^{+}(N, t)-\bar{\rho}(N-1, t)v^{+}(N-1, t)]
$$

$$
-[\bar{\rho}(N+1, t)v^{-} (N+1, t)-\bar{\rho}(N, t)v^{-}(N, t)]=0
\eqno(A1)
$$

\noindent where $v^{+}(N, t)$ represents the fraction of phase points in the $N^{th}$ hyperplane leaving to the $(N+1)^{th}$ plane per unit time, $v^{-}(N, t)$ represents the fraction of phase points in the $N^{th}$ hyperplane leaving to the $(N-1)^{th}$ plane per unit time, etc.  Now, using a finite difference representation of a derivative, we can write

$$
\frac{\partial (\bar{\rho}v^{+})}{\partial N}\cong 
\frac{\bar{\rho} (N, t)v^{+}(N, t)-\bar{\rho}(N-1, t)v^{+}(N-1, t)}{N-(N-1)}\eqno(A2)
$$

\noindent Thus, Eq.(A1) becomes in the limit of large $N$

$$
\frac{\partial \bar{\rho}}{\partial t} + \frac{\partial (\bar{\rho} v^{+})}{\partial N} - \frac{\partial (\bar{\rho} v^{-})}{\partial N}=0 \eqno(A3)
$$

\noindent or, 

$$
\frac{\partial \bar{\rho}}{\partial t}+\frac{\partial (\bar{\rho} v)}{\partial N}=0
\eqno(A4)
$$

\noindent where 

$$
v \equiv v^{+}-v^{-}\eqno(A5)
$$

\end{document}